\documentstyle[12pt]{article}
\begin{document}
\setlength{\unitlength}{1mm}
{\hfill  JINR E2-94-117, March 1994 } \\
\begin{center}
{\Large\bf  Exact Solution Of 2D Poincar\'e Gravity Coupled To
Fermion Matter}
\end{center}
\begin{center}
{\large\bf Sergey Solodukhin$^{\ast}$}
\end{center}
\begin{center}
{\bf Bogoliubov Laboratory of Theoretical Physics,
Joint Institute for Nuclear Research,
Head Post Office, P.O.Box 79, Moscow, Russia}
\end{center}
\vspace*{2cm}
\begin{abstract}
   The 2D model of gravity with zweibeins $e^{a}$ and the Lorentz
connection one-form $\omega^{a}_{\ b}$ as independent gravitational variables
coupled to 2d massless Dirac matter is considered. It is shown that the
classical
equations of motion are exactly integrated in the case of chiral fermions.
\vspace*{2cm}
\end{abstract}
\vskip 4cm
\noindent $^{ \ast}$ e-mail: solodukhin@main1.jinr.dubna.su

\newpage

{\bf 1.}
The numerous recent attempts to formulate the
theory of gravity in the framework of a consistent gauge approach resulted in
constructing the gauge gravity models for the de Sitter and Poincar\'e groups
(for a review
see, e.g., [1] ). The independent variables are now vielbeins
$e^{a}= e^{a}_{\mu}dx^{\mu}$ and Lorentz connection one-form
$\omega^{a}_{\ b} = \omega^{a}_{\ b,\mu} dx^{\mu}$.
These methods being applied in two dimensions, give us  an dynamical
 description of 2D
 gravity.
  It was argued that investigation of simple two-dimensional model leads to
a better understanding of four-dimensional gravity and its quantization [2].
It was shown in [2] that the Lagrangian $L= \gamma R^{2} + \beta
T^{2} + \lambda $ is the most general one quadratic in curvature
$R$ and torsion
$T$, and containing a cosmological constant $\lambda$. The classical equations
of motion for this type of two-dimensional gravity were analyzed in conformal
gauge
 [3] and in light cone gauge [4] and
their exact integrability was demonstrated. The various aspects of quantization
of the model were recently considered in [5]. In ref.[6] was shown that the
formulation
of the model on the language of differential forms is very useful. This allows
to
find exactly the solution of vacuum gravitational equations using an
appropriate
(and rather natural) coordinates on the 2D space-time. The resulting metric
can be written in the Schwarzschild-like form and describes asymptotically
de Sitter black hole configuration [6]. Using this method in [7] one
proves the integrability of the general 2D Poincar\'e gauge gravity with
Lagrangian being an
arbitrary (not necessary quadratic) function of curvature and torsion and
demonstrates that the field equations is again of the black hole type.

The coupling with matter in general case breaks this exact integrability.
One exceptional case noted in [6,7] is the 2D Yang-Mills field.
In this letter we consider the coupling the 2D Poincar\'e gauge gravity with
2D massless Dirac fermions and show that the resulting field equations are
exactly integrated by means the method of ref.[6].

\bigskip

 {\bf 2}.  We begin with brief description the Poincar\'e gauge gravity and
 Dirac spinors in two dimensions
\footnotemark\
\addtocounter{footnote}{0}\footnotetext{The exhausted introduction to 2D Dirac
spinors one can find in [7]}. In this letter we follow notations of
 paper [6].
 The 2D gauge gravity is described in terms of zweibeins
$e^{a} = e^{a}_{\mu} dz^{\mu}, a=0,1 $ (the 2D metric on the surface $M^{2}$
has the form $g_{\mu \nu}=e^{a}_{\mu} e^{b}_{\nu} \eta_{ab} $,
 $\eta_{ab}= diag (+1,-1)$) and Lorentz
connection one-form $\omega^{a}_{\ b} = \omega \varepsilon^{a}_{\ b}, \  \omega
=
\omega_{\mu} dz^{\mu} \  (\varepsilon_{ab} =- \varepsilon_{ba}, \
\varepsilon_{01}=1)$.
The curvature and torsion two-forms
are:
\begin{equation}
R=d\omega, \ \ \ T^{a}=de^{a} + \varepsilon^{a}_{\ b}
\omega \wedge e^{b}
\end{equation}
With respect to the Lorentz connection $\omega$ one can define the covariant
derivative $\nabla$ which acts on the Lorentz vector $A^a$ as follows

$$
\nabla A^a := dA^a + \varepsilon^a_{\ b} \omega \wedge A^b
$$

\bigskip

The Dirac matrices $\gamma^a, \ a=0,1$ in two dimensions satisfy the relations:
\begin{equation}
\gamma^a\gamma^b=\eta^{ab}-\varepsilon^{ab} \gamma_5
\end{equation}
where $\gamma_5=\gamma^0\gamma^1, \ (\gamma_5)^2=1$.
The following identities are also useful:
\begin{equation}
\gamma^a \gamma_5+\gamma_5 \gamma^a=0
\end{equation}
and
\begin{equation}
\gamma^a \gamma_5=\varepsilon^a_{\ b} \gamma^b
\end{equation}
In further consideration we use explicit realization of $\gamma$-matrices:
\begin{equation}
\gamma^0= \left( \begin{array} {ll}
                   0 & 1 \\
                   1 & 0
                   \end{array} \right) \ \
\gamma^1= \left( \begin{array} {ll}
                   0 & 1 \\
                   -1 & 0
                   \end{array} \right) \ \
\gamma_5= \left( \begin{array} {ll}
                   -1 & 0 \\
                   0 & 1
                   \end{array} \right) \ \
\end{equation}
The Dirac spinors in two dimensions have two complex components:
\begin{equation}
 \Psi = \left( \begin{array}{c}
             \psi_{1} \\ \psi_2
             \end{array} \right)
\end{equation}
and under local Lorentz rotation (on angle $\Omega$) transform as follows
\begin{equation}
\Psi \rightarrow \Psi' =S \Psi, \ \ \bar{\Psi} \rightarrow \bar{\Psi}'
=\bar{\Psi} S^{-1}
\end{equation}
where the Dirac conjugated spinor is defined as $\bar{\Psi}=\Psi^+ \gamma^0$.
Matrix $S$ realizing the spinor representation of 2D Lorentz group is given by
\begin{equation}
S= \cosh [{\Omega \over 2}] - \gamma_5 \sinh [{\Omega \over 2}]
\end{equation}
One can see that components $\psi_1$ and $\psi_2$ transform independently:
\begin{equation}
\psi'_1 =e^{\Omega \over 2} \psi_{1} , \ \
\psi'_2 =e^{-{\Omega \over 2}} \psi_{2}
\end{equation}
This means that left(right)-chiral spinors defined as
\begin{equation}
\gamma_5 \Psi = \mp \Psi
\end{equation}
give us the irreducible representations of the Lorentz group.

\bigskip

It is useful to define the covariant spinor derivative $\nabla$ as
differential operator acting on the field $\Psi$ considered as zero-form with
values
in two-dimensional complex spinor space:
\begin{equation}
\nabla \Psi := d \Psi +{1 \over 2} \omega \gamma_5 \Psi , \ \
\nabla \bar{\Psi} := d \bar{\Psi} -{1 \over 2} \omega  \bar{\Psi} \gamma_5
\end{equation}
This definition means that operator $\nabla$ acts on spinor bilinear
combinations, such as $\bar{\Psi} \Psi, \ \bar{\Psi} \gamma^a \Psi , \
\bar{\Psi} \gamma^{[a} \gamma^{b]} \Psi$, as usual covariant derivative on
Lorentz scalar, vector and bivector correspondingly.
One can see from (11) that spinor covariant derivative $\nabla$ acts on
components
of spinor field (6) as follows

$$
\nabla \psi_1 =d\psi_1-{1 \over 2} \omega \psi_1 \ ; \ \nabla \psi_2 =
d \psi_2 +{ 1 \over 2} \omega \psi_2
$$

\bigskip

{\bf 3}. The dynamics of 2D gravitational  $( e^{a}, \omega)$ and
fermion ($\Psi$) variables is determined by the action:
\begin{equation}
S=S_{gr}+S_{fer},
\end{equation}
where
\begin{equation}
S_{gr}= \int\limits_{M^{2}}^{} {\alpha \over 2} \ast T^{a} \wedge T^{a} +
{1 \over 2} \ast R \wedge R - {\lambda^{} \over 4} \varepsilon_{ab} e^{a}
\wedge e^{b}
\end{equation}
is standard action of 2D Poincar\'e gauge gravity quadratic in
curvature and torsion;  $\ast$ is the Hodge dualization and $\alpha,\lambda$
are arbitrary constants.

The action for 2D Dirac fermions in terms of differential forms can be written
as
follows:
\begin{equation}
S_{fer}= \int_{}^{}{\imath \over 2} \varepsilon_{ab} e^a \wedge (\bar{\Psi}
\gamma^b
\nabla \Psi - \nabla \bar{\Psi} \gamma^b \Psi )
\end{equation}
Notice that in this letter we consider only the massless fermions. One can see
that due to identity (3) the Lorentz connection $\omega$ is dropped out
from expression (14) and really one can use the usual external derivative
$d$ instead of $\nabla$ in (14).

\bigskip

Instead the curvature $R$ and torsion $T^a$ two-forms let us consider the dual
zero-forms $\rho=\ast R, \ q^a= \ast T^a$.

The variation of (12) with respect to the Lorentz connection $\omega$ and
zweibeins $e^a$ gives the  following equations
\begin{equation}
d \rho=-\alpha q^a \varepsilon_{ab} e^b
\end{equation}
\begin{equation}
 \nabla q^a =-{ 1\over 2\alpha} \Phi (\rho, q^2) \varepsilon^{a}_{\ b} e^b +
J^a,
\end{equation}
where  $q^{2}=q^{a}q^{b} \eta_{ab} $. In (16) the following
notation was introduced:
$\Phi (q^2,\rho)=\rho^{2} +\alpha q^{2}- \lambda $.
The matter one-form $J^a$ takes the form:
\begin{equation}
J^a=-{\imath \over 2} \varepsilon^a_{\ b} (\bar{\Psi} \gamma^b \nabla \Psi -
\nabla \bar{\Psi}  \gamma^b \Psi )
\end{equation}
It should be noted that  $J^a=J^a_\mu dx^\mu$ is related with matter
energy-momentum tensor: $T_{\mu\nu}= {1 \over 2} (\varepsilon_\mu^{\ \alpha}
J^a_\alpha e^a_\nu + \varepsilon_{\nu}^{\ \alpha} J^a_\alpha e^a_\mu)$.

Variation of action (12) with respect to fermion field $\Psi$ gives equation:
\begin{equation}
(e^a \varepsilon_{ab} \gamma^b) \wedge (\nabla \Psi) ={ 1\over 2} T^a
\varepsilon_{ab}
\gamma^b \Psi
\end{equation}

\bigskip

{}From (1) we obtain that
\begin{equation}
\omega = \check{\omega} -q_a e^a
\end{equation}
where $\check{\omega}$ is torsionless part of the Lorentz connection:
\begin{equation}
de^a+ \varepsilon^a_{\ b} \check{\omega} \wedge e^b=0
\end{equation}
Using (19) and identity (4) the equation (18) can be rewritten as follows
\begin{equation}
(e^a \varepsilon_{ab} \gamma^b) \wedge (d \Psi + { 1 \over 2} \check{\omega}
\gamma_5 \Psi)=0
\end{equation}
i.e. the torsion is dropped in the Dirac equation.
Taking the Hodge dualization of (21) one can transform (21) to more standard
form of
the Dirac equation:

$$
\gamma^\mu(\partial_\mu+{1 \over 2} \check{\omega}_\mu \gamma_5)\Psi=0
$$

where $\gamma^\mu=e^\mu_a \gamma^a$.

\bigskip

Using the Dirac equation (21) one can show that one-forms $J^a$ (17)satisfy
following identities:
\begin{equation}
J_a \wedge e^a=0 , \ \ \  \varepsilon_{ab} J^a \wedge e^b =0
\end{equation}
Really (22) are consequences of invariance action (14) under local
Lorentz and conformal transformations correspondingly [8].

\bigskip

The  components of the spinor field (6) can be written as $\psi_i=e^{\chi_i}$,
$i=1,2$, where $\chi_1=\beta+ \imath v, \ \chi_2=\gamma+\imath u$ are complex
fields.
Then the one-forms $J^a$ (17) take the form
\begin{equation}
J^0=[e^{2\gamma}du-e^{2\beta}dv], \ \
J^1=[e^{2\gamma}du+e^{2\beta}dv],
\end{equation}
while the Dirac equation (18) reads
\begin{eqnarray}
&&(e^0-e^1)\wedge (\imath du+d\gamma+ {1 \over 2}\omega)e^\gamma=
{1 \over 2}(T^0-T^1)e^\gamma
\nonumber \\
&&(e^0+e^1)\wedge (\imath du+d\gamma- {1 \over 2}\omega)e^\beta=
{1 \over 2}(T^0+T^1)e^\beta
\end{eqnarray}

\bigskip

{\bf 4.} Assume that the orthonormal basis $\{e^a \}$ takes the
conformal-Lorentz
form:
\begin{equation}
e^a=e^\sigma (n^a d \tau - \varepsilon^{a}_{\ b} n^b dx)
\end{equation}
where $n^a, a=0,1$ is unite Lorentz vector, $n^2=n^an_a= \pm 1$. By means of
diffeomorphism transformations in two dimensions arbitrary basis $\{ e^a \}$
always can be transformed to the form (25).

The corresponding metric $ds^2= \eta_{ab} e^a_\mu e^b_\nu dx^\mu dx^\nu$ takes
the
conformally flat form:

$$
ds^2=n^2 e^{2\sigma} (d \tau^2-dx^2)
$$

By means the identity
\begin{equation}
A \wedge e^a=\varepsilon^a_{\ b}e^b \wedge (\ast A),
\end{equation}
where $A$ is arbitrary one-form, we get for differential of (25):
\begin{equation}
de^a=\varepsilon^a_{\ b}(- \ast (d\sigma)+n^\alpha \varepsilon_{\alpha\beta}
dn^\beta) \wedge e^b.
\end{equation}
Inserting (27) into (20) we obtain for $\check{\omega}$:
\begin{equation}
\check{\omega}=\ast(d\sigma)-n^a \varepsilon_{ab}dn^b.
\end{equation}
Assuming for definiteness that $n^2=1$, components $n^a$ can be written as
$n^0=\cosh \theta, \ n^1= \sinh \theta$. So we have that $n^a \varepsilon_{ab}
dn^b=d \theta$. Under local Lorentz rotation on angle $\Omega$ variable
$\theta$ transforms as $\theta \rightarrow \theta-\Omega$. So the last term in
(28)
is pure gauge part of the Lorentz connection.

Substituting the expression (28) into the Dirac equation (21) and using
identities
(4) and (26) we get
\begin{equation}
(e^a \varepsilon_{ab} \gamma^b)\wedge (d+{1 \over 2}d\sigma -{1 \over 2}
d\theta
\gamma_5 ) \Psi
\end{equation}
or in spinor components (6):
\begin{eqnarray}
&&(e^0+e^1)\wedge (d+{1 \over 2} d\sigma \psi_1 +{ 1 \over 2} d\theta )\psi_1=0
\nonumber \\
&&(e^0-e^1)\wedge (d+{1 \over 2} d\sigma \psi_1 -{ 1 \over 2} d\theta )\psi_2=0
\end{eqnarray}
For basis (25) we have

$$
(e^0 \mp e^1)=e^{(\sigma \mp \theta)} (d \tau \pm dx)
$$

Taking this into account, the equations (30) are easily solved and we obtain
for the
spinor field:
\begin{equation}
 \Psi = e^{-{\sigma \over 2}} \left( \begin{array}{c}
             e^{-{\theta \over 2}} e^{\imath v(x^-)}p(x^-) \\
             e^{{\theta \over 2}} e^{\imath u(x^+)}f(x^+) \\
             \end{array} \right)
\end{equation}
where $v,p$ and $u,f$ are arbitrary functions  of the light-cone coordinates
$x^-=\tau -x$ and $x^+=\tau+x$ correspondingly.

Thus the Dirac equation (18),(21), taken separately, is exactly solved in
the conformal-Lorentz gauge (25) and general solution takes the form
(31). However, now one must put the (31) in the gravitational eqs.(15),(16) and
find the joint solutions of the coupled gravity-Dirac system.

{\bf 5.} As in vacuum case [6], there are two types of solutions of
eqs.(15)-(18).
The first one is characterized by that the torsion squared is zero,
 $q^2 \equiv 0$. One can see from eqs.(15)-(18) that it is possible only in the
case
when torsion is identically zero: $q^a \equiv 0, \ a=0,1$, the space-time has
constant curvature
 \footnotemark\
\addtocounter{footnote}{0}\footnotetext{Note that only if $\lambda \geq 0$
there exists the constant curvature solution.} :
 $\rho^2=\lambda$,
and the one-forms (17) vanish:
$J^a \equiv 0, \ a=0,1$.

If zweibeins are taken in the form (25) the constant curvature condition:
$\ast(d \omega)=\rho=const$, gives us the equation for conformal factor
$\sigma$:
$\ast d \ast (d \sigma)=\rho=const$, which is equivalent to the Liouville
equation:
\begin{equation}
2 \partial_- \partial_+ \sigma={\rho \over 2} e^{2\sigma},
\end{equation}
where $\rho=\pm \sqrt{\lambda}$. The general solution of the Liouville equation
is well-known.
By means the coordinate changing it can be transformed to the form:

$$
2\sigma=- \ln (1-{\rho \over 4} x^+ x^-)^2 .
$$

Correspondingly, we have for metric

$$
ds^2={dx^+ dx^- \over (1-{\rho \over 4} x^+ x^-)^2}
$$

and  for Lorentz connection (28):

$$
\omega={{\rho \over 4} \over 1-{\rho \over 4} x^+x^-}(x^+dx^- - x^-
dx^+)-d\theta
$$

The one possible solution for the Dirac field is trivial, $\Psi=0, \
(e^\gamma=e^\beta=0)$. The non-trivial $\Psi$ with vanishing forms $J^a$ (23)
is given
by (31) where $u$ and $v$ are constant functions:
\begin{equation}
 \Psi = (1-{\rho \over 4} x^+ x^-)^{1 \over 2} \left( \begin{array}{c}
             e^{-{\theta \over 2}} e^{\imath v}p(x^-) \\
             e^{{\theta \over 2}} e^{\imath u}f(x^+) \\
             \end{array} \right)
\end{equation}

\bigskip

{\bf 6.} Let us now assume that $q^2 \neq 0$ identically on 2D space-time. We
begin the analysis with the case when $J^a=0, \ a=0,1$. Then the gravitational
field equations (15), (16) completely decouples from the Dirac equation (18).
One sees
from (23) that $J^a$ vanish if $e^\gamma, \ e^\beta$ are zero or/end the
imaginary
part of $\chi_i, \ u$ and $v$, are constant functions. The gravitational
equations reduce
to the vacuum case. The general vacuum solution was obtained in [6]
(for more accurate definitions see [9]). It is essential that one uses
the variable $\rho$ as one of the space-time coordinates. Introducing $\phi$ as
additional, orthogonal to $\rho$, coordinate, we can write the vacuum solution
for the zweibeins:
\begin{equation}
e^a=q^a e^{-{\rho \over \alpha}} d \phi -{ 1 \over \alpha q^2} \varepsilon
^a_{\ b} q^b
d\rho
\end{equation}
and for the Lorentz connection:
\begin{equation}
\omega=-{ 1 \over q^2} q^a \varepsilon_{ab} dq^b -{ \alpha \over 2} (q^2)'_\rho
e^{-{\rho \over \alpha}} d \phi,
\end{equation}
where $q^2$ is known function of $\rho$:
\begin{equation}
q^2(\rho)=-{ 1 \over \alpha}(\rho+\alpha)^2+\Lambda +\epsilon e^{\rho \over
\alpha},
\end{equation}
where $\Lambda=\lambda / \alpha -\alpha$, $\epsilon$ is the integrating
constant.

The corresponding metric
\begin{equation}
ds^2=q^2 e^{-2\rho \over \alpha} d \phi^2 - {1 \over \alpha^2 q^2} d\rho^2
\end{equation}
was shown to describe the asymptotically de Sitter black hole configuration
with ADM mass proportional to $\epsilon$. The zeros of $q^2$ are points of
the horizons [6].

It is worth observing that (34) takes the form (25) if we identify:
$n^a={q^a \over q}$, $ e^\sigma=qe^{-{\rho \over \alpha}}$,
$\tau=\phi$, $ x= \int_{}^{\rho}{e^{\rho' \over \alpha} \over \alpha q^2
(\rho')}
d\rho'$. For definiteness we assume that $q^2 >0$, then  $q \equiv \sqrt{q^2}$.

Indeed, in coordinates $(\phi, x)$ the metric (37) is conformally flat:
\begin{equation}
ds^2=q^2(\rho) e^{-2{\rho \over \alpha}}(d \phi^2 -dx^2)
\end{equation}
where $\rho$ can be, in principle, expressed as function of $x$.
Note again that the first term in (35) is pure gauge: ${1 \over q^2} q^a
\varepsilon_{ab}
dq^b=d \theta$.

Since the solution of the Dirac equation (18), (21) for zweibeins
taken in the form (25) is already known (31), we obtain the following
expression
for the fermion field:
\begin{equation}
 \Psi =q^{-1/2} e^{\rho \over 2\alpha} \left( \begin{array}{c}
             e^{-{\theta \over 2}} e^{\imath v}p(x^-) \\
             e^{{\theta \over 2}} e^{\imath u}f(x^+) \\
             \end{array} \right)
\end{equation}
where $u$ and $v$ are constants, and $x^{\mp}=\phi \mp x$. We see that
$\Psi$ (39) divergences at points where $q^2$ has zeros. Remember that these
points are regular horizons of the vacuum metric (37), (38). Nerveless, nothing
singular
happens at these points since the energy-momentum tensor for the spinor
configuration (39)
is identically zero. The fermion field  $\Psi$ also divergences at the point
$e^{-{\rho \over \alpha}}=0$,
where the black hole singularity is located (see [6]), while it tends to zero,
$\Psi \rightarrow
0$, if $e^{-{\rho \over \alpha}} \rightarrow \infty$.

\bigskip

{\bf 7.} Let us assume that $q^2 \neq 0$ identically on 2D space-time.
The fermion action (14) is invariant under (global) chiral ($\gamma_5$)
transformations: $\Psi \rightarrow \Psi'=exp[\mu \gamma_5] \Psi$.
Therefore for simplicity we may restrict ourselves by consideration
 only the  fermions of fixed chirality:
$$
\gamma_5 \Psi = \Psi.
$$

In this case the fermion field has only one non-zero component:
$\psi_1=0, \ \psi_2=e^\chi$,
where $\chi=\gamma+ \imath u$ is complex field.

Then only the first of equations  (24) is non-trivial. It gives us, in
particular, that
$du \sim (e^0-e^1)$. In Lorentz invariant form it can be written as follows:
\begin{equation}
q_a e^a -q^a \varepsilon_{ab} e^b = B du,
\end{equation}
where $B$ is still unknown scalar function.
As it is seen from (9), only the real part of $\chi$ transforms under
Lorentz group: $\gamma \rightarrow \gamma -{\Omega \over 2}$. So the imaginary
part, $u$,
 is Lorentz scalar.

\bigskip

One can see  from (15) and (40) that variables $\rho$ and $u$ can be naturally
chosen as coordinates on 2D space-time. Then basis of one-forms $e^a$ is
expressed in terms of ($d\rho, \ du$):
\begin{equation}
e^a ={q^a \over q^2} (-{ d\rho \over \alpha} +B du) -{ 1 \over \alpha q^2}
\varepsilon^a_{\ b}
q^b d\rho
\end{equation}
The metric $ds^2= \eta_{ab} e^a_\mu e^b_\nu  dx^\mu dx^\nu$ correspondingly
takes the form:
\begin{equation}
ds^2={ 1\over q^2} (B du - {d\rho \over \alpha})^2-{1 \over \alpha^2 q^2}
d\rho^2
\end{equation}

\bigskip

In terms of the field $\chi=\gamma+ \imath u$ the one-form $J^a$ (23) has the
components:

$$
J^0=J^1=e^{2\gamma}du
$$

It is convenient to introduce the one-form

$$
J={2 \over q^0 + q^1} e^{2\gamma} du
$$

Assuming for definiteness that $q^2>0$ let us introduce variable $\theta$:
$q^0=q \cosh \theta, \ q^1= q \sinh \theta, \ q \equiv \sqrt{q^2}$.
Then we have for $J$:

$$
J={ 2 \over q} e^{2\gamma-\theta} du
$$

Under local Lorentz rotation on angle $\Omega$ variable $\theta$  transforms
as: $\theta \rightarrow \theta -\Omega$. So the combination $(2\gamma-\theta)$
is really
Lorentz invariant.

\bigskip

Multiplying eq.(16) on $q^a$ and $q^b \varepsilon_{ba}$ separately we obtain
equations:
\begin{equation}
dq^2= {\Phi \over \alpha^2}d\rho+q^2J
\end{equation}
\begin{equation}
\omega+d\theta=-{\Phi \over 2\alpha q^2} q_a e^a +{ 1 \over 2} J
\end{equation}
where we used that ${ 1 \over q^2} q^a \varepsilon_{ab}dq^b =d\theta$.
The Lorentz connection $\omega$ with respect to Lorentz rotations transforms as
$\omega \rightarrow \omega+d\Omega$. So that $(\omega+d\theta)$ is again
Lorentz
invariant. The eq.(43) gives us $q^2$ as function of $\rho$ and
$u$, while  (44) is equation on the Lorentz connection $\omega$.
The (43) is equivalent to
\begin{equation}
\partial_\rho q^2={\Phi \over \alpha^2} (\rho, q^2), \ \ \partial_u q^2 =2q
e^{2\gamma-\theta}
\end{equation}
It follows from the first eq.(45) that $q^2$ as function of $\rho$ has
the same form as in vacuum case [6] (see eq.(36)).
However, the $\epsilon$ now is
 function of $u$, $\epsilon=\epsilon(u)$, which is found from
the second eq.(45). Taking into account that $\partial_u q^2=\partial_u
\epsilon
e^{\rho / \alpha}$ we get
\begin{equation}
\partial_u \epsilon =2q e^{-\rho / \alpha} e^{ 2\gamma - \theta}\end{equation}
Since the left hand side of eq.(46) is function of only variable $u$ we
obtain that $(2\gamma-\theta)$ must have the following form:
\begin{equation}
2\gamma-\theta=- \ln q +{\rho \over \alpha} +2 \ln f(u),
\end{equation}
where $f(u)$ is an function of variable $u$ related with $\epsilon (u)$
by means of equation:
\begin{equation}
\partial_u \epsilon=2f^2(u).
\end{equation}

\bigskip

Acting now  by external differential $d$ on both sides of eq.(40) we obtain:
\begin{equation}
B'_{\rho} d\rho \wedge du=( {\Phi \over \alpha} -q^2) V +
J_a \wedge e^a -\varepsilon_{ab} J^a \wedge e^b
\end{equation}
{}From (41) we have $d\rho \wedge du={\alpha \over B} q^2 V$. Then using
 (22) and (45) the eq.(49) gives us
 the equation on function $B$:
\begin{equation}
{B'_\rho \over B}= { 1 \over q^2} \partial_\rho q^2 -{ 1 \over \alpha}
\end{equation}
{}From this we finally find:
\begin{equation}
B=B_0(u)q^2 e^{-{\rho \over \alpha}}
\end{equation}
where $B_0$ is an arbitrary function of $u$.
Now inserting (51)  into eq.(44) we obtain the expression for the Lorentz
connection:
$$
\omega+d\theta= -{\alpha \over 2} \partial_\rho q^2 e^{-{\rho \over \alpha}}
B_0(u) du  +{1 \over 2 q^2} \partial_\rho q^2 d\rho + { 1 \over q^2}
e^{\rho \over \alpha} f(u) du.
$$
Taking into account eq.(48) we finally obtain
\begin{equation}
\omega+d\theta= -{\alpha \over 2} \partial_\rho q^2 e^{-{\rho \over \alpha}}
B_0(u) du  +{1 \over 2 } d (\ln q^2 )
\end{equation}
It should be noted that modulo exact forms this expression for $\omega$
takes the same form as in vacuum case [6] (see (35)).

Now it is easy to check the self-consistency condition: $\ast (d\omega)=\rho$.
Really this procedure is the same as in vacuum case.

  Let us again consider the Dirac equation  (24).
It is easy to see  from (41) that

$$
e^0-e^1=B_0 q e^{-{\rho \over \alpha}} e^{-\theta} du
$$
Inserting this and eq.(52) into the first equation (24) we obtain
\begin{equation}
B_0 e^{-{\rho \over \alpha}} du \wedge (d\gamma - { 1 \over 2} d\theta +{ 1
\over 2q}
\partial_\rho q d\rho)=- { 1 \over 2} V
\end{equation}
Using the obtained expressions for $\gamma$ (47) and eq.(48) we obtain that
(53)  holds identically.

   This completes the proof of exact integrability of equations (15)-(18).
The complete solution is given by expression

$$
e^a ={q^a \over q^2} (-{ d\rho \over \alpha} + q^2 e^{-{\rho \over \alpha}} B_0
du) -{ 1 \over \alpha q^2} \varepsilon^a_{\ b}
q^b d\rho
$$
for zweibeins; expression (52) for the Lorentz connection $\omega$ and
\begin{equation}
\Psi = q^{-1/2} e^{\theta \over 2} e^{\rho \over 2\alpha}
              \left( \begin{array}{c}
         0 \\ e^{i u} f(u)
         \end{array}  \right)
\end{equation}
for the chiral fermion field.
The $q^2$ is known function of $\rho$ and $u$:

$$
q^2(\rho)=-{ 1 \over \alpha}(\rho+\alpha)^2+\Lambda +\epsilon (u) e^{\rho \over
\alpha},
$$
where

$$
\epsilon (u)= 2 \int_{}^{u} f^2(u')du'
$$

Note that up to
this moment everything was Lorentz invariant.
As result, the general solution depends on arbitrary field
$\theta$ that is reflection of underlying Lorentz symmetry.
Now one can fix the gauge, say
$\theta=0$ (see [9]).

   The solution also depends on arbitrary function $f(u)$ which is not
determined from the field equations and is found from initial conditions for
fermion field.

  In the case when fermions of both chiralities present the eqs.(15)-(18) can
  be integrated in the same manner taking the imaginary parts, $u$ and $v$, of
the
  spinor components (see (31)) as light-cone coordinates.
  However, solution takes more complicated form.

\bigskip

{\bf 8}. The sense of found solution becomes more transparent if we
consider $\delta$-like impulse of fermion matter:
\begin{equation}
f^2(u)={ E \over 2} \delta(u-u_0), \ E>0
\end{equation}
Then equation (48) is easily solved:
\begin{equation}
\epsilon (u)= \epsilon_0 +E \theta (u-u_0)
\end{equation}
where $\theta(x)$ is step function.
In regions $u<u_0$ and $u>u_0$ taken separately the  function $\epsilon (u)$
is constant and one can consider here new variable $v$:
\begin{equation}
v=u-\int_{}^{\rho}{ 2e^{\rho' \over \alpha} \over \alpha q^2(\rho')} d\rho'
\end{equation}
Then in coordinates $(u,v)$ the metric (42) takes the vacuum conformally flat
form (38):
\begin{equation}
ds^2=q^2 e^{-2\rho \over \alpha}dudv
\end{equation}
For $u<u_0$ we have vacuum black hole solution (34)-(38) with mass
$\epsilon=\epsilon_0$.  The fermion impulse with energy $E$ falls into this
space-time
at $u=u_0$ along $v$-direction. In result, for $u>u_0$ we again obtain vacuum
black hole solution but with mass $\epsilon=\epsilon_0+E$.

It was shown in [6] that the space-time structure of the vacuum solution
(34)-(38)
essentially depends on value of the constant $\epsilon$. The falling of the
fermion matter leads to
re-construction of initial vacuum accordingly to new value of $\epsilon$. It
should be noted that
in this aspect the found solution is similar to that of the 2D dilaton gravity
coupled with scalar (conformal) matter [10]. However, there are some essential
differences.
The flat space-time is one of solutions in 2D dilaton gravity. The falling of
the scalar matter
 into the flat space-time leads to  formation of the black hole. In the
case under consideration there is no such a solution describing the black hole
formation
from regular space-time (in our case it is the de Sitter one) due to fermion
matter. The "bare" vacuum black hole configuration is necessary. The reason is
that the vacuum constant curvature solution is not obtained from the black hole
one
(34)-(38) for an value of integrating constant $\epsilon$, i.e. these solutions
are not parametrically connected
\footnotemark\
\addtocounter{footnote}{0}\footnotetext{Really this situation is typical for 2D
gravity described by action polynomial in curvature [11].}. Instead, in 2D
dilaton gravity
[12] the flat space-time is obtained as zero mass black hole solution.

\bigskip

{\bf 9.}  In conclusion, we studied the 2D Poincar\'e  gauge gravity
coupled to 2D massless Dirac fermions and showed that the classical
equations are exactly integrated. As in vacuum case, there are two types of
solutions.
The solution of the first type is space-time of constant curvature
$(\rho^2=\lambda)$ and zero torsion, $q^a=0, \ a=0,1$. The corresponding
fermion field can take trivial $(\Psi =0)$ and non-trivial configurations.
The solution of second type is characterized by that torsion is not
identically zero. The space-time is of the  black hole type with mass
dependent on in-coming fermion matter energy.

\bigskip

I would like to thank Prof.F.W.Hehl and Dr.Yu.N.Obukhov for kind
hospitality at University of Cologne. This work was supported in part by the
grant 94-02-03665-a of Russian Fund of Fundamental Investigations.

\end{document}